\documentclass{llncs}
\usepackage[longnamesfirst, authoryear]{natbib}
\usepackage{amsmath}
\usepackage{amsfonts}
\usepackage{orcidlink}
\DeclareRobustCommand\widebar[1]{\mathop{\overline{#1}}}

\begin{document}

\title{Author-Unification: \\ Name-, Institution-, and Career-Sharing Co-authors}
\titlerunning{Author-Unification (AUA)}
\authorrunning{Vanessa Wirth}

\author{
Vanessa Wirth\orcidlink{0000-0001-8295-3021}\thanks{The authors contributed equally to this work} \and Vanessa Wirth\orcidlink{0009-0000-3595-1250}$^*$
}

\institute{
Frequent-Authornames-University Erlangen-Nürnberg (FAU)\\
\email{vanessa.wirth@fau.de}\thanks{Mails will be automatically forwarded to the other Vanessa Wirth anyway.}
}

\maketitle

\small

\begin{figure}[t]
	\centering
	\includegraphics[width=0.6\linewidth]{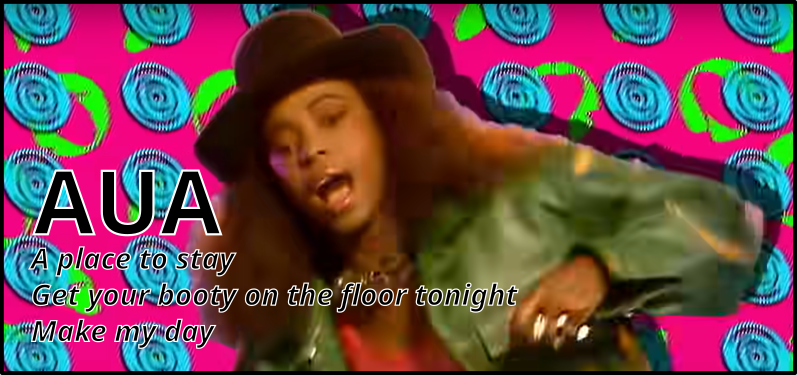}
	\caption{In 1989, the Belgian act \textit{Technotronic} already stressed the importance of Author Unification (AUA). In their song \textit{Pump up the Jam}, AUA is referred to as the future 'place to stay'. It encourages researchers to search for this future place to stay and we are the first in doing so. The interested reader can listen to the pronunciation of \textit{AUA} at \url{https://www.youtube.com/watch?v=9EcjWd-O4jI}(00:39s).}
	\label{fig:teaser}
\end{figure}

\begin{abstract}
  In this work, we investigate the phenomenon of \textit{\textbf{A}uthor-\textbf{U}nific\textbf{A}tion} (AUA), which describes the high structural similarity of two co-authoring engineers that share the same forename, surname, institution, and academic career without being related by blood. So far, prior work has only explored similar surnames and institutions. On top of that, we elaborate on the additional author similarity of sharing the same academic career as a Ph.D. candidate with the same starting day and month included in the university contract. We show that our work outperforms previous state-of-the-art investigations, among others by providing a higher \textbf{S}tructural \textbf{S}imilarity \textbf{I}ndex \textbf{M}easure (SSIM) of the letters in our names and in our institution. Lastly, we prove the duality of our identities through a qualitative evaluation.
\end{abstract}

\section{Introduction and Related Work}
\label{sec:intro}
As concluded by~\cite{goodman}, sharing a co-authorship with authors of equal surnames is a promising method to achieve more success in academia whilst avoiding the \textit{"et al"} penalty and alphabetical discrimination. 
Hence, it is desirable to collaborate with researchers of a similar name. \\
However, with respect to the current state-of-the-art, a major challenge still persists: 
It is very cumbersome to establish contact with authors of the same surname if they are not blood-related. 
As the likelihood of researchers being busy is remarkably high, a natural consequence is that they do not respond often to messages from unknown contacts.
Hence, it is of significant interest to increase their attention with respect to fruitful future collaborations.  This can be achieved by enforcing social pressure, e.g. through physical presence or, in other words, through constant physical positioning in front of a researcher's office. 
However, only a few rare scenarios, in which the co-authors shared the same campus, such as~\cite{chen}, \cite{rosen}, and \cite{otto} exist so far. 
This is probably because their university was not as great as ours. \\
Furthermore, even though sharing the same surname avoids alphabetical discrimination, a major research gap stems from the fact that all contributing authors of similar works can still be distinguished from each other by their forenames or institution. To the best of our knowledge, the only work, in which the authors shared the same forename, surname, and university, is proposed by~\cite{otto} very recently. However, the authors forgot to address the benefits of having the same forename, surname, and university. \\
In summary, the advantages of sharing co-authorship among authors of similar names are still heavily under-explored. 
We aim to bridge this research gap by investigating the phenomenon of two co-authors sharing the exact same name and academic institution. 
To the best of our knowledge, among similar works in economics (\citep{goodman,reinhart}), psychology (\cite{sue}), and statistics (\cite{otto}), we are the first authors in the field of engineering\footnote{We leave the reviewers to the longstanding debate, whether computer science is a discipline of science, mathematics or engineering. Have fun!}. 
On top of that, we explore how the concept of career-sharing, i.e. sharing the same starting day and month of pursuing an academic career as a Ph.D. candidate, leads to ultimate success. 
We refer to the phenomenon of sharing all the above-mentioned author properties, such that the authors become indistinguishable, as \textit{Author-Unification} (AUA), in accordance with the definition given by the Belgian act \textit{Technotronic} in \autoref{fig:teaser}. We describe the characteristics of AUA in the following sections and highlight their superiority.

\section{The phenomenon of Author-unification}
\label{sec:unification}
The extension from so-called \textit{Surname-Sharing} authors \cite{goodman} to an \textit{Author-Unification} is very novel and there is a high risk that this corner case was not even considered in some academic software systems or websites. To foster the early detection of a possible Author-Unification, the following sections address its characteristics. Furthermore, we show their potential of leading to a higher success in an academic career.

\subsection{Name-sharing of Co-authorships}
When a publication is cited, it is oftentimes referred to as \textit{Author et al.} within a particular text passage. 
Unfortunately, in most cases in which the author is able to use the citation function properly, the term \textit{Author} denotes only the first author. To avoid discriminating against all remaining co-authors,~\cite{goodman} proposed to leverage co-authorships with common surnames instead. 
However, a remaining problem persists in the bibliography section, in which the literature is usually sorted alphabetically by the full name of the first author. 
On top of that, with different forenames, the necessity remains to list all authors separately and, thus, occupy additional page space, which could be potentially used to address important research results, e.g. that this paper has to be read by everyone\footnote{This statement is already a proof itself that we can leverage additional page space. Furthermore, this paper has to be read by everyone.}.  
Instead, the phenomenon of name-sharing co-authorships, describing authors of the exact same name, presents the opportunity to summarize all authors into a unified name. 
In this way, it is possible to avoid discrimination and multiple word repetitions in the bibliography section.
We also encourage authors to have a surname with a small number of letters as proposed by~\cite{wirth}. This can be achieved by being lucky, marrying a person with a short surname or by introducing an alter ego, for example, \textit{Kanye East}.
With this method, it is possible to save up to exactly half a page by summarizing citations and literature items into a unified and short name.

\begin{figure}[t]
	\centering
	\includegraphics[width=\linewidth]{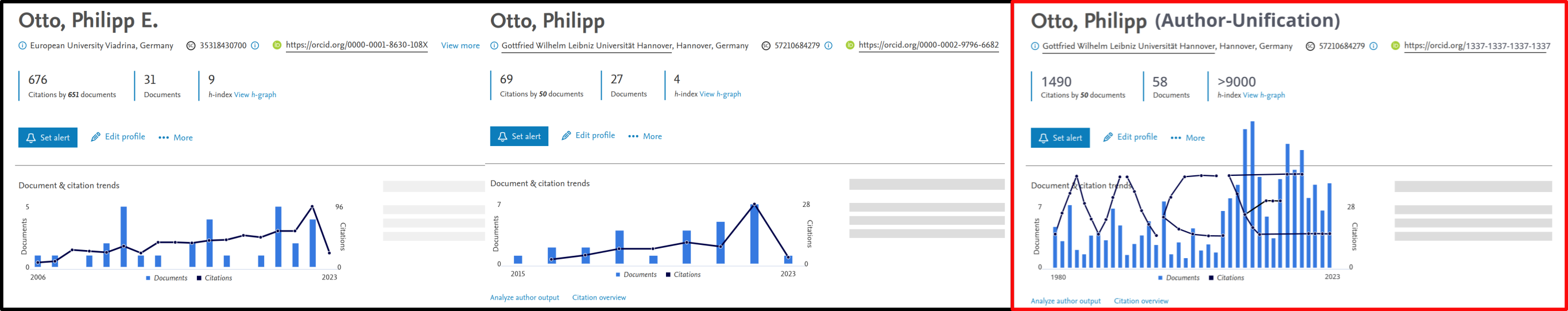}
	\caption{Left: Overview of the individual academic career of both Philipp Ottos from~\cite{otto} (extracted from \url{https://www.scopus.com}). Right: Bundled academic success if both authors would have unified into one for the greater good. We observe a hint to the acronym NICE, referring to \textbf{N}ame-, \textbf{I}nstitution-, and \textbf{C}areer-sharing \textbf{E}ngineers. Apparently, the statistics, generated by the AI of \cite{opencheat}, could also predict the fundamental principles of our work.}
	\label{fig:statistics}
\end{figure}

\subsection{Co-authorships within the same institution}
Beyond the scope of each individual publication, a unified author name can be extended to personal publication records, social media platforms, or any general type of registration form on a website. 
In this way, the pain of managing digital, personal content and the conglomeration of software and website bugs that come along with it, can be distributed either evenly or unevenly (we recommend the latter). 
Most importantly, a common digital presence increases international awareness and popularity by a factor of X (for eXtreme) and strengthens the academic career along with it. In \autoref{fig:statistics}, we show how the academic success of~\cite{otto} could have looked like if both authors agreed to an AUA. \\
However, to establish the above-mentioned opportunities, a shared professional mail from a common academic institution is required. 
We argue that with the same name, the necessity of a personal institutional mail address becomes irrelevant either way because all name-sharing individuals will certainly receive the same messages altogether. 
Based on empirical studies, there is a probability of probably 76.5\% of mistakenly sending a letter or mail or, even more frequently, a fax to the wrong forename and surname-sharing person. 
Ground-truth data, indicating whether the recipient was mistaken for another, was collected among the confused (in German: \textit{"ver-wirth"}) co-workers. 
For more information on the quality of this study, we encourage the reviewers to write a message to the author of this section (cf. author contribution on the last page). \\
To summarize, by sharing the same name and institution, we see the possibility to share the same institutional mail address, as we observe an increasing non-discriminability, non-discrimination, and non-discretion within the institution itself.
Non-discriminability leads to further advantages such as uninvited participation in research events, exchanging balance cards for the cafeteria, and more intra-institutional nett-working as the number of known co-workers who know one of the name-sharing authors inevitably will connect with the respective others at some point in time.

\subsection{Co-authorships of similar academic career}
Lastly, we discuss the benefits of sharing an academic career by pursuing a Ph.D. starting from the same day and month in time. 
First, we argue that authors, which are in similar stages of their academic careers also share similar goals, i.e. they just want to get their \textit{shit} done. 
As sharing the same name and institution inevitably leads to a connection with the future co-author, we observe a strong and natural opportunity to achieve those goals together by collaborating on several publications, i.e. getting their \textit{sheets} done. 
For example, in the work of~\cite{wirth} it becomes clear that the authors have a similar recurrence frequency of publication deadlines due to their similar Ph.D. careers. 
Thus, the strong urge to publish a paper at the famous conference of \textit{the ACH Special Interest Group on Harry Quakeproof Bovik (SIGBOVIK)} was equally present among both authors. 
Another advantage of a similar career is that, after a paper has been accepted, only one celebration cake needs to be baked and shared among the co-workers.

\section{Results}
\label{sec:results}
In this section, we provide qualitative and quantitative results of our work. 
First, we compare our work with current state-of-the-art on a quantitative benchmark.
In \autoref{table:quantitative}, we evaluate four different high-quality metrics, which are described in the next paragraphs.

\begin{table}[] 
\centering
\begin{tabular}{ c||c|c|c|c|| } 
& NSA $\uparrow$ & GEIL $\downarrow$  & SSIM $\uparrow$ & ACDC $\uparrow$ \\ 
\hline
\hline
\cite{otto} & \textbf{2.00} & \textbf{1.00} & 52.00 & 2.00 \\
\cite{chen} & 0.00 & \textbf{1.00} & 2.71 & 1.67 \\
\cite{goodman} & 0.00 & 1.50 & 0.73 & 8.54 \\
\hline
\textbf{Ours} & \textbf{2.00} & \textbf{1.00} & \textbf{96.00} & \textbf{$\boldsymbol \infty$} \\

\end{tabular}
\caption{Quantitative results of ours and previous work. The best results are highlighted in bold.}
\label{table:quantitative}
\end{table}

\paragraph{\textbf{NSA}.} The \textbf{N}umber of name-\textbf{S}haring \textbf{A}uthors metric describes the number of name-sharing authors. It is an indicator for a fruitful academic career as more authors of the same name are able to bundle their individual success into significantly greater success. As depicted in \autoref{table:quantitative}, our work is on par with the current state-of-the-art of \cite{otto} while outperforming all remaining works.

\paragraph{\textbf{GEIL}.} The \textbf{GE}nder \textbf{I}mbalance \textbf{L}evel encodes the gender diversity among all co-authors by
evaluating the following formula on the super-set of multivariate genders $\widehat{\mathcal{G}} = \{ \mathcal{G}_1, ..., \mathcal{G}_{\lvert \widehat{\mathcal{G}} \rvert } \}$, 
in which each subset $\mathcal{G}_*$ with $\lvert \mathcal{G}_* \rvert \geq 1$ contains all authors plus one random person over the world, which define themselves as a specific gender $*$:
\begin{equation*}
       \textbf{GEIL} =  \frac{1}{2\lvert \widehat{\mathcal{G}}-1 \rvert} \sum_{\mathcal{G}_1 \in  \widehat{\mathcal{G}}}\sum_{\substack{\mathcal{G}_2 \in \widehat{\mathcal{G}}, \\ \mathcal{G}_1\neq \mathcal{G}_2}} \lVert {\lvert \mathcal{G}_1 \rvert} - { \lvert \mathcal{G}_2 \rvert}\rVert_1,
\end{equation*}
For reasons of simplicity, we only include the genders \textit{female}, \textit{male}, and \textit{diverse} in our super-set $\widehat{\mathcal{G}}$. However, we note that the GEIL can be extended to other genders as well. In summary, a GEIL closer to zero indicates a better balance among the genders.
In view of the presented results, we conclude that all works are still far away from the optimum.

\paragraph{\textbf{SSIM}.} The SSIM measures the similarity between author names and their institutions and is given by:
\begin{align}
    \textbf{SSIM} &= \left[ \exp{\left( \text{A} \right)} +  \exp{\left( \text{U} \right)} \right] \cdot \underset{u \in \mathcal{U}}{ \max} \left( \textbf{L}(u) \right) \\ 
    \text{A} &= -\frac{1}{\lvert \mathcal{A} \rvert -1} \left( \sum_{a_1 \in  \mathcal{A}}\sum_{\substack{a_2 \in \mathcal{A}, \\ a_2\neq a_1}} \mathbf{I}(f_1,f_2) + \mathbf{I}(s_1,s_2) \right) \\
    \text{U} &= - \frac{1}{\lvert \mathcal{U} \rvert -1}  \left( \sum_{u_1 \in  \mathcal{U}}\sum_{\substack{u_2 \in \mathcal{U}, \\ u_2\neq u_1}} \mathbf{I}(u_1,u_2) \right)
\end{align}
The function \textbf{L} returns the number of letters of an institution $u \in\mathcal{U}$. 
The terms A and U measure the dissimilarity between each author $a_i \in \mathcal{A}$ (with forename $f_i$ and surname $s_i$), and their institution $u_i$, respectively. 
More specifically, the indicator function \textbf{I}$(x_1,x_2)$ returns 1 if at least one letter of $x_1$ differs from $x_2$, and 0 otherwise. 
In summary, \autoref{table:quantitative} shows that our work outperforms the current state-of-the-art by a large margin.
As the formula was generated by~\cite{zero2hero} and is way too complex for an analytic conclusion, we figure, our outstanding performance is achieved due to the exact same name and same institution, which has the longest name of them all.
\paragraph{\textbf{ACDC}.} The \textbf{A}cademic \textbf{C}areer-\textbf{D}octoral candidate \textbf{C}orrelation  is an indicator for the temporal overlap of academic careers as a Ph.D. and measures the correlation of the first starting day $d_i$ and month $m_i$ of the university contract between each pair of authors:
\begin{equation*}
       \textbf{ACDC} = \left( \frac{\sum_{i=1}^{\lvert \mathcal{A} \rvert} \sum_{j=1}^{\lvert \mathcal{A} \rvert} (d_i - \widebar{\textbf{d}})(d_j - \widebar{\textbf{d}})} {\sum_{i=1}^{\lvert \mathcal{A} \rvert} d_i - \widebar{\textbf{d}}} \right) 
       + \left( \frac{\sum_{i=1}^{\lvert \mathcal{A} \rvert} \sum_{j=1}^{\lvert \mathcal{A} \rvert} (m_i - \widebar{\textbf{m}})(m_j - \widebar{\textbf{m}})} {\sum_{i=1}^{\lvert \mathcal{A} \rvert} m_i - \widebar{\textbf{m}}} \right)
\end{equation*}
We denote the mean starting day and mean starting month across all authors as $\widebar{\textbf{d}}$ and $\widebar{\textbf{m}}$, respectively. 
To compare our work with the state-of-the-art, we searched for the respective Ph.D. starting date and month to allow a fair comparison among all the authors on the same academic level. 
In case we could not find the relevant information, we chose a random value determined by an objective third-party unit, i.e. a cat controlled by a laser beam pointed at our keyboard while listening to \textit{Back in Black}. 
Note that for reasons of privacy, we will not publish our findings. The results are depicted in \autoref{table:quantitative} and were shocking us \textit{all night long, yeah}. 
When computing the ACDC for our work, we experienced a remarkably high correlation of infinity caused by an unforeseen division by zero of our program\footnote{As program execution is always deterministic, we expect similar results for the other metrics in case of zero-divisions. }. 
In this regard, it becomes clear that our work surpasses all the others.
\begin{figure}[t]
	\centering
	\includegraphics[width=0.7\linewidth]{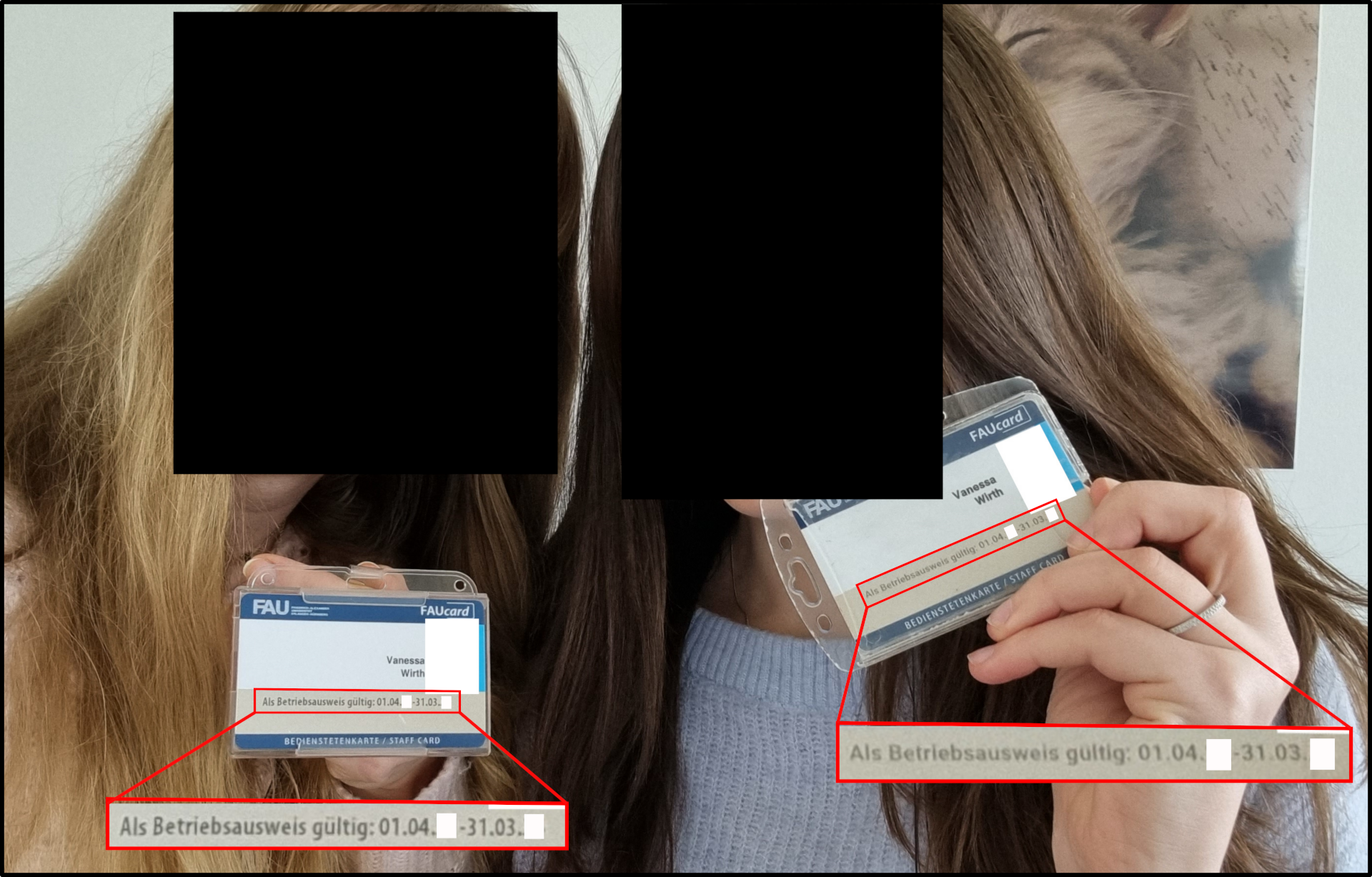}
	\caption{Qualitative proof of dual identity of the authors. We apologize for potential headaches caused by the umlauts on our staff card.}
	\label{fig:qualitative}
\end{figure}

\paragraph{\textbf{Qualitative Evaluation}.} To proof the duality of the authors identity, we provide an anonymized qualitative identity measure in \autoref{fig:qualitative}. We believe it is destiny that our first working day is the 1st of April.

\section{Conclusion}
In summary, our work outperforms them all. Nevertheless, with respect to the GEIL a lot of further research still has to be done. A possible future direction is to find co-authors with gender-neutral names and/or similar appearance. We believe, our work is the stepping stone, but no stumbling stone, for further research in the field of Author-Unification.

\paragraph{\textbf{Acknowledgements and Author Contribution}.} 
We thank \textit{et al}  and \textit{Bernhard Egger} for their partially valuable feedback. The abstract and \autoref{sec:intro} were written by Vanessa Wirth. Remaining sections, i.e. \autoref{sec:unification} and \autoref{sec:results}, were written by Vanessa Wirth.

\bibliographystyle{plainnat}
\bibliography{bibliography}

\end{document}